\documentclass[12pt,a4paper]{article}
 \usepackage{epsf}
\begin{document}
 \title{The bound on the mass of the new gauge boson $Z^{\prime}$ from the process
 $\mu\longrightarrow 3e$}
 \author{Chongxing Yue$^{(a,b)}$,  Guoli Liu$^{b}$, Jiantao Li$^{b}$ \\
 {\small a: CCAST (World
  Laboratory) P.O. BOX 8730. B.J. 100080 P.R. China} \\
  {\small b: Department of Physics, Henan Normal University,
  Xingxiang  453002. P.R.China}
 \thanks{This work is supported by the National Natural Science
  Foundation of China, the Excellent Youth Foundation of Henan Scientific
  Committee; and Foundation of Henan Educational Committee.}
 \thanks{E-mail: cxyue@pbulic.xxptt.ha.cn} }
 \date{\today}
 \maketitle
 \begin{abstract}
 \hspace{5mm}The new gauge boson $Z^{\prime}$ predicted by the strong top
 dynamical symmetry breaking models has significant contributions
 to the lepton flavor changing process $\mu\longrightarrow 3e $. We consider
 the bound on the mass of the new gauge boson $Z^{\prime}$ from the experimental
 value of the branching ratio $Br(\mu\longrightarrow 3e)$ in the framework of
 topcolor assisted technicolor models. We find that the precision
 experimental value of $Br(\mu\longrightarrow 3e)$ gives a
 severe bound on the $Z^{\prime}$ mass $M_{Z^{\prime}}$. For $k_1 \leq 1$,
 $M_{Z^{\prime}}$ must be larger than $1.64$ $TeV$.
 \end {abstract}
 \vspace{1.0cm}

 \newpage
    The generation of a large fermion mass such as $m_t=175$ $GeV$ is
 a difficult problem in the theories of dynamical electroweak symmetry
 (EWS) breaking. Technicolor (TC) \cite{b1} with an extended
 technicolor (ETC) \cite{b2} can naturally break the EWS to give
 rise to the weak gauge boson masses and also generate the masses
 of ordinary quarks and leptons. However, ETC models can not explain
 the top quark's large mass without running afoul of the experimental
 constraints from the parameter T and the $Z\longrightarrow
 b\overline{b}$ branching ratio $R_b$ \cite{b3}. Top quark
 condensation models \cite{b4} try to identify all of the EWS
 breaking with the formation of a dynamical top quark mass, but
 this requires a very large scale $\Lambda\sim 10^{15}GeV$ for the
 new dynamics and significant fine tuning.

    The large mass of the top quark suggests that it may play a
 special role in the dynamics of the EWS breaking and flavor
 symmetry breaking. The topcolor assisted technicolor (TC2) models
 \cite{b5}, the top see-saw models \cite{b6} and the flavor
 universal coloron models \cite{b7} are three of such examples.
  These models predict the existence of colored gauge bosons
 (topgluons, colorons), color-singlet gauge boson ($Z^{\prime}$),
 Pseudo Goldstone bosons (technipions and top-pions), and heavy fermions.
 These new particles can be seen as characteristics of these models.
 Studying the effects of these new particles in various process
 will be of particular interest.

   The new strong or flavor interactions may exist at relatively
 low scales and may play an integral part in either EWS breaking or
 fermion mass generation. Thus, it is interesting to study current
 experimental bounds on the mass of the corresponding gauge bosons.
 Ref.[8] gives the limits on the mass of the new gauge bosons $Z^{\prime}$
 via studying its corrections to the precisely measured electroweak
 quantities at $LEP$ and its effects on bijet production and single top
 production at Tevatron. In this letter, we will discuss the bounds on
 the mass of the new gauge boson $Z^{\prime}$ from the lepton flavor changing
 process $\mu\longrightarrow 3e$ in the framework of TC2 models.
 Our results show that the precision experimental value of $B_r(\mu
 \longrightarrow 3e)$ gives a severe bound on the $Z^{\prime}$ mass
 $M_{Z^{\prime}}$. For the parameter $k_1\leq 1$, $M_{Z^{\prime}}$ must be larger
 than the $1.64$ $TeV$. This is consistent with the limit obtained in Ref.[8].

   In the standard model (SM), because of the strong GIM suppression,
 the tree-level flavor changing neutral currents are absent. Lepton flavor
 changing processes are strongly suppressed by powers of small neutrino masses.
 This opens the possibility of using the corresponding flavor changing
 process to probe new physics, whose effects may include
 appreciable violation of neutral flavor conservation already
 probed by present high energy colliders. The underlying
 interactions in the strong top dynamical symmetry breaking models
 (such as TC2 models and top see-saw models) are non-universal and
 therefore do not possess a GIM mechanism. This is an essential
 feature of this kind of models due to the need to single out the
 top quark for condensation. When the non-universal interactions
 are written in the mass eigenstates, it may lead to the flavor
 change coupling vertices of the new gauge bosons, such as $Z^{\prime}tc$,
 $Z^{\prime}\mu e $, $Z^{\prime}\mu \tau$. Thus, the new gauge boson $Z^{\prime}$
 may have significant contributions to some lepton flavor changing processes.
 The lepton flavor changing processes may have severe bound on the
 mass $M_{Z^{\prime}}$ of the $Z^{\prime}$. Thus, we can give the
 bound on the mass $M_{Z^{\prime}}$ of the $Z^{\prime}$ via
 discussing its contributions to these lepton flavor changing
 processes.

 At present, the branching ratios of the lepton flavor changing
 $\mu$ decay processes, such as $\mu \rightarrow 3e$,
 $\mu\rightarrow e \gamma$ and $\mu \rightarrow e \gamma\gamma$,
 have been measured precisely. There are severe bounds on these
 decay processes, i.e. $Br(\mu \rightarrow 3e)\leq 10^{-12}$,
 $Br(\mu\rightarrow e \gamma)\leq 4.9\times 10^{-11}$,
 $Br(\mu \rightarrow e \gamma\gamma)\leq 10^{-10}$\cite{b9}.
 The $Z^\prime$ have no contributions to the processes
 $\mu\rightarrow e \gamma$ and $\mu \rightarrow e \gamma\gamma$,
 the experimental value of the branching ratios can not give any
 bound on the $M_{Z^{\prime}}$. The new gauge boson $Z^{\prime}$
 have contributions to the lepton flavor changing $\tau$ decay
 processes, such as $\tau\rightarrow 3e$, $\tau\rightarrow
 e\mu\mu$, etc. Compared to the process $\mu \rightarrow 3e$,
 however, the experimental bounds on these processes are weaker.
 The branching ratios are of order $10^{-6}$\cite{b9}. Thus, in this
 paper, we will concentrate on the bound on the $M_{Z^{\prime}}$
 from the lepton flavor changing process $\mu \rightarrow 3e$.

     In TC2 models, the ETC interactions have contributions to all
 quark and lepton masses, while the mass of the top quark is mainly
 generated by the topcolor interactions, and EWSB is driven by
 technicolor or a Higgs sector. To maintain electroweak symmetry
 between top and bottom quarks and yet not generate $m_b \simeq
 m_t$, the topcolor gauge group is usually taken to be a
 strongly coupled $SU(3)\otimes U(1)$. The $U(1)$ provides the
 difference that causes only top quarks to condense. At the $\Lambda\sim 1
 TeV$, the dynamics of a general TC2 model involves the following
 structure \cite{b5,b10}:
 \begin{equation}
   SU(3)_1  \otimes SU(3)_2 \otimes U(1)_{y_1} \otimes U(1)_{y_2}
 \times SU(2)_L \longrightarrow SU(3)_{QCD} \otimes U(1)_{EM}
 \end{equation}
 where $SU(3)_1\otimes U(1)_{y_1}$ ($SU(3)_2 \otimes U(1){y_2}$)
 generally couples preferentially to the third (first and second )
 generations. The $U(1)_{y_i}$ are just strongly rescaled versions
 of electroweak $U(1)_y$. This breaking scenario gives rise to the
 topcolor gauge bosons including the color-octet coloron
 $B_{\mu}^A$ and color-singlet extra $U(1)$ gauge boson $Z^\prime$. The
 coupling of the new gauge boson $Z^\prime$ and $B^A_{\mu}$ to ordinary
 fermions can be written as :
 \begin{equation}
 {\cal{L}}_{Z^\prime}=g_1 cot\theta^\prime Z^\prime \cdot J_{Z^\prime},
 \hspace{8mm}
 {\cal{L}}_B=g_3 cot\theta B^A \cdot J^A_B
 \end{equation}
 where $g_3$($g_1$) is the QCD($U(1)_y$) coupling constant at the
 scale $\Lambda_{TC}$, $\theta$ and $\theta^\prime$ are the mixing
 angles. To obtain the top quark direction for condensation, we
 have $cot\theta \gg 1$ and $cot\theta^\prime \gg 1$.
   Integrating out the heavy bosons $Z^\prime$ and $B^A_{\mu}$, the couplings
 (Eq.(2)) give the effective low energy four fermion interactions,
  which can be written as :
 \begin{equation}
 {\cal{L}}_{eff,Z^\prime}=-\frac{2\pi k_1}{M_{Z^\prime}^2}J_{Z^\prime}\cdot J_{Z^\prime},
 \hspace{8mm}
 {\cal{L}}_{eff,B}=-\frac{2\pi k}{M^2_B}J^A_B \cdot J^A_B
 \end{equation}
 where $M_{Z^\prime}$ and $M_B$ are the masses of the new gauge boson $Z^\prime$
 and $B^A_{\mu}$, respectively.
 $k_1$ and $k$ are coupling constants which can be written as $k_1=g_1^2
 {cot^2\theta^\prime}/4\pi$, $k=g^2_3cot^2\theta/4\pi$. In general,
 the currents $J_{Z^\prime}$ and $J_B$
 involve all three generations of fermions:
 \begin{equation}
 J_Z^\prime=J_{Z^\prime,1}+J_{Z^\prime,2}+J_{Z^\prime,3},\hspace{8mm}
 J_B=J_{B,1}+J_{B,2}+J_{B,3}
 \end{equation}
 For the first and second generations, the currents are (in weak
 eigenbasis):
 \begin{eqnarray}
 J_{Z^\prime,1}^{\mu}&=&-\tan^2{\theta^\prime}\left(\frac{1}{6}
 \overline{u_L}\gamma^{\mu}u_L
 +\frac{1}{6}\overline{d_L}\gamma_{\mu}d_L +\frac{2}{3}\overline{u_R}
 \gamma^{\mu}u_R \right. \\ \nonumber
 &&-\frac{1}{3}\overline{d_R}\gamma^{\mu}d_R-\frac{1}{2}e_L
 \gamma^{\mu}e_L-\frac{1}{2}\overline{\nu_{eL}}\gamma^{\mu}\nu_{eL}
 -\overline{e_R}\gamma^{\mu}e_R \left. \right)
 \end{eqnarray}
 \begin{eqnarray}
 J_{Z^\prime,2}^{\mu}&=&-\tan^2{\theta^\prime}\left(\frac{1}{6}
 \overline{c_L}\gamma^{\mu}c_L
 +\frac{1}{6}\overline{s_L}\gamma^{\mu}s_L+\frac{2}{3}\overline{c_R}
 \gamma^{\mu}c_R \right. \\ \nonumber
 &&-\frac{1}{3}\overline{s_R}\gamma^{\mu}\overline{s_R}-\frac{1}{2}
 \mu_{L}\gamma^{\mu}\mu_L-\frac{1}{2}\nu_{\mu L}\gamma^{\mu}\nu_{\mu
 L}-\mu_{R}\gamma^{\mu}\mu_{R} \left. \right)
 \end{eqnarray}
 \begin{equation}
 J^{\mu}_{B,1}=-\tan^2{\theta} (\overline{u}\gamma^{\mu}\frac{\lambda^A}{2}u
 +\overline{d}\gamma^{\mu} \frac{\lambda^A}{2}d)
 \end{equation}
 \begin{equation}
 J_{B,2}^{\mu}=-\tan^2{\theta} (\overline{c} \gamma^{\mu}
 \frac{\lambda^A}{2}c+\overline{s}\gamma^{\mu}
 \frac{\lambda^A}{2}s)
 \end{equation}
 where $\lambda^A$ is a Gell-Man matrix acting on color indices.
 From Eq.(7) and Eq.(8), we can see that the gauge bosons
 $B^A_{\mu}$ have no contributions to the lepton flavor changing process
 $\mu \longrightarrow 3e$. The precision experimental value of
 $Br(\mu \longrightarrow 3e)$ can not give any bound on the mass $M_B$
 of the color-octet coloron $B^A_{\mu}$.

    For TC2 models, the underlying interactions, topcolor
 interactions, are non-universal and therefore do not possess a GIM
 mechanism. When the non-universal interactions are written in the
 mass eigenstates, it results in the flavor changing coupling
 vertices. After rotation to the mass eigenstates, Eq.(3) generates
 four fermion interactions leading to the flavor changing coupling
 vertices. For the lepton flavor changing process
 $\mu\longrightarrow 3e$, the relative effective Lagrangian can be
 written as:
 \begin{equation}
 {\cal{L}}_{eff}^\prime=\frac{\pi
 k_1\tan^4{\theta^\prime}}{2M_{Z^\prime}^2}[k_L(\mu_L \gamma^{\mu}e_{L})(e_L
 \gamma_{\mu}e_L)+2k_R(\mu_{R}\gamma^{\mu}e_{R})(e_{R}\gamma_{\mu}e_R)]
 \end{equation}
 where $k_L$ and $k_R$ are the flavor mixing factors. In the
 following estimation, we will assume $|k_L|=|k_R| \simeq \lambda$
 \cite{b10,b11}, which $\lambda$ is the Wolfenstein parameter\cite{b12}.

   Comparing the contributions of the gauge boson $Z^\prime$ to the process
 $\mu\longrightarrow 3e$ to that of ordinary muon decay
 $\mu\longrightarrow e\nu\overline{\nu}$, which proceeds via the
 electroweak gauge boson $W$ exchange, gives the branching ratio
 $Br(\mu\longrightarrow 3e)$ arising from the $Z^\prime$ exchange:
 \begin{equation}
 Br(\mu\longrightarrow 3e)=\frac{\Gamma(\mu\longrightarrow 3e)
 }{\Gamma(\mu\longrightarrow e\nu\overline{\nu})}
 =\frac{5\alpha^2_e
 S_W^4 M_W^4}{64C_W^8 k_1^2 M_{Z^\prime}^4}A
 \end{equation}
 with $A=k_L^2+4k_R^2$, $S_W=\sin\theta_W$ and $C_W=\cos\theta_W$,
 which $\theta_W$ is the weinberg angle, $\alpha_e$ is the
 electromagnetic coupling constant. In our estimation, we will
 take $\lambda=0.22$, $\alpha_e=1/128.9$, $M_W=80.41$ $GeV$ \cite{b13}.
  Using the experimental value $Br(\mu\longrightarrow 3e)\leq
 10^{-12}$, we can give the bound on the mass $M_{Z^\prime}$ from Eq.(10).
 Our results are presented in Fig1. From Fig1, we can see that
 the lower bound on the mass $M_{Z^\prime}$ increases with decreasing the
 value of $k_1$. Considering the requirement of vacuum tilting and
 the constraints from Z-pole physics and $U(1)$ triviality, there is
 the region of coupling constant parameter space which is $k\simeq
 2$, $k_1\leq 1$ for TC2 models \cite{b10,b14}. If we take
 $k_1=0.2$, then we have $M_{Z^\prime}\geq 3.68$ $TeV$.

 From Eq.(10), we can see that the bound on $M_{Z^\prime}$ is
 sensitive to the values of the flavor mixing factors. To see the effect
 of the flavor mixing factors on the bound, we plot bound on the
 mass $M_{Z^\prime}$ as a function of the parameter $\lambda$ in
 Fig2 for $k_1=1$. From Fig2, we can see that the lower bound on
 the mass $M_{Z^\prime}$ increases with increasing the value of
 $\lambda$. For $\lambda \geq 0.21$, there must be $M_{Z^\prime}\geq
 1.6TeV$.

 In this paper, we have discussed the bound on the mass $M_{Z^\prime}$
 of the new gauge boson $Z^\prime$ from the experimental value of
 the branching ratio $Br(\mu\rightarrow 3e)$ in the framework of
 TC2 models. Our results shows that the precision experimental
 value of $Br(\mu\rightarrow 3e)$ gives a stringent bound on
 $M_{Z^\prime}$. The mass $M_{Z^\prime}$ must be larger than
 $1.64TeV$ for $k_1\leq 1.0$ and $\lambda=0.22$.
 \newpage
 \vskip 2.0cm
 \begin{center}
 {\bf Figure Captions}
 \end{center}
 \begin{description}
 \item[Fig.1:]The lower bound on the mass $M_{Z^\prime}$ as a function of
 the parameter $k_1$. The horizontal line is the bound on the
 parameter $k_1$.
 \item[Fig.2:]The lower bound on the mass $M_{Z^\prime}$ as a
 function of the parameter $|k_L|=|k_R|=\lambda$ for $k_1=1$.
 \end{description}
 
 \newpage
 \begin{figure}
 \begin{center}
 \begin{picture}(300,80)
 \put(-85,-75){\epsfxsize150mm\epsfbox{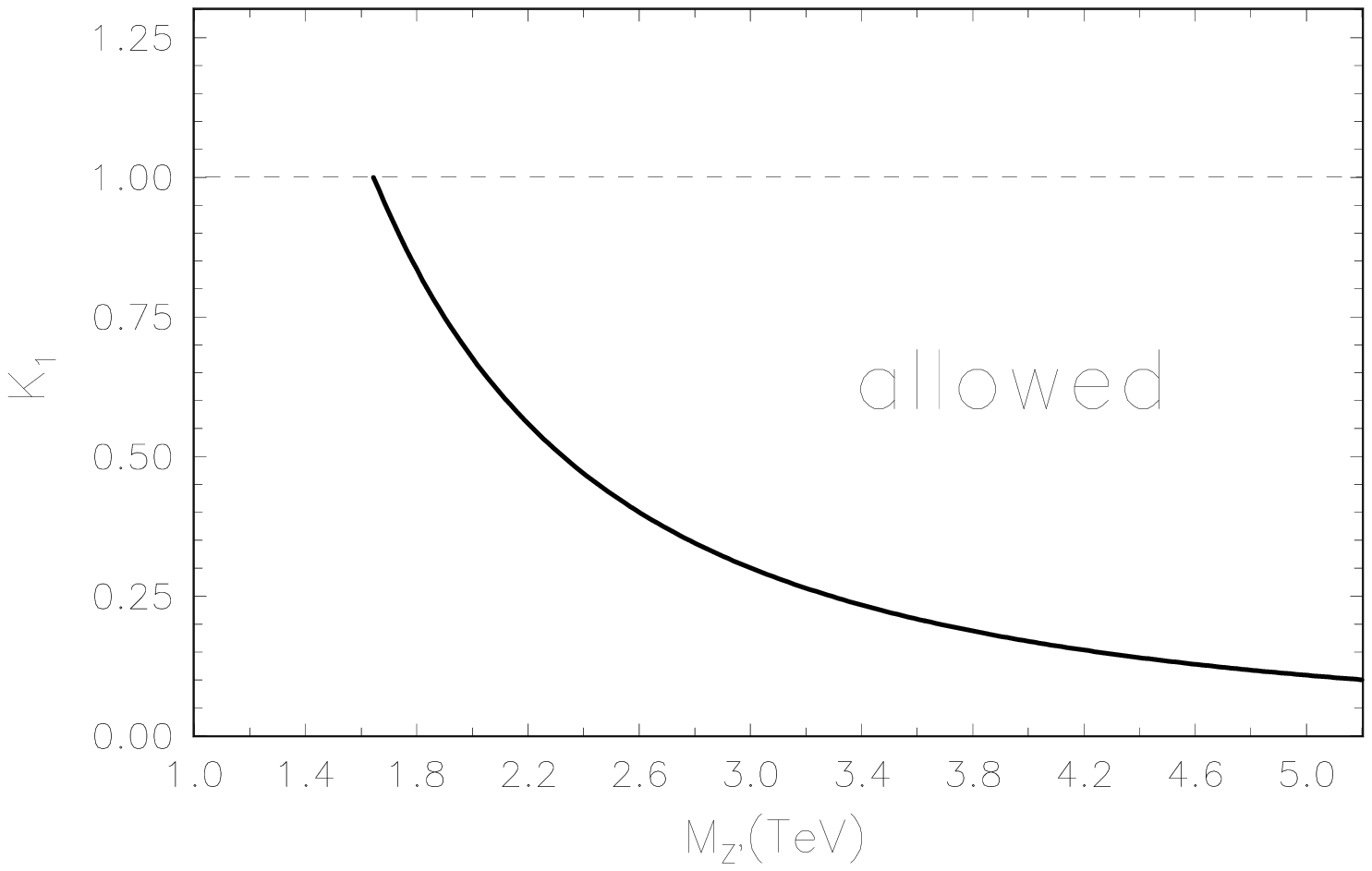}}
 \put(120,-95){\huge Fig.1}
 \end{picture}
 \end{center}
 \end{figure}

 \begin{figure}
 \begin{center}
 \begin{picture}(300,80)
 \put(-85,-75){\epsfxsize150mm\epsfbox{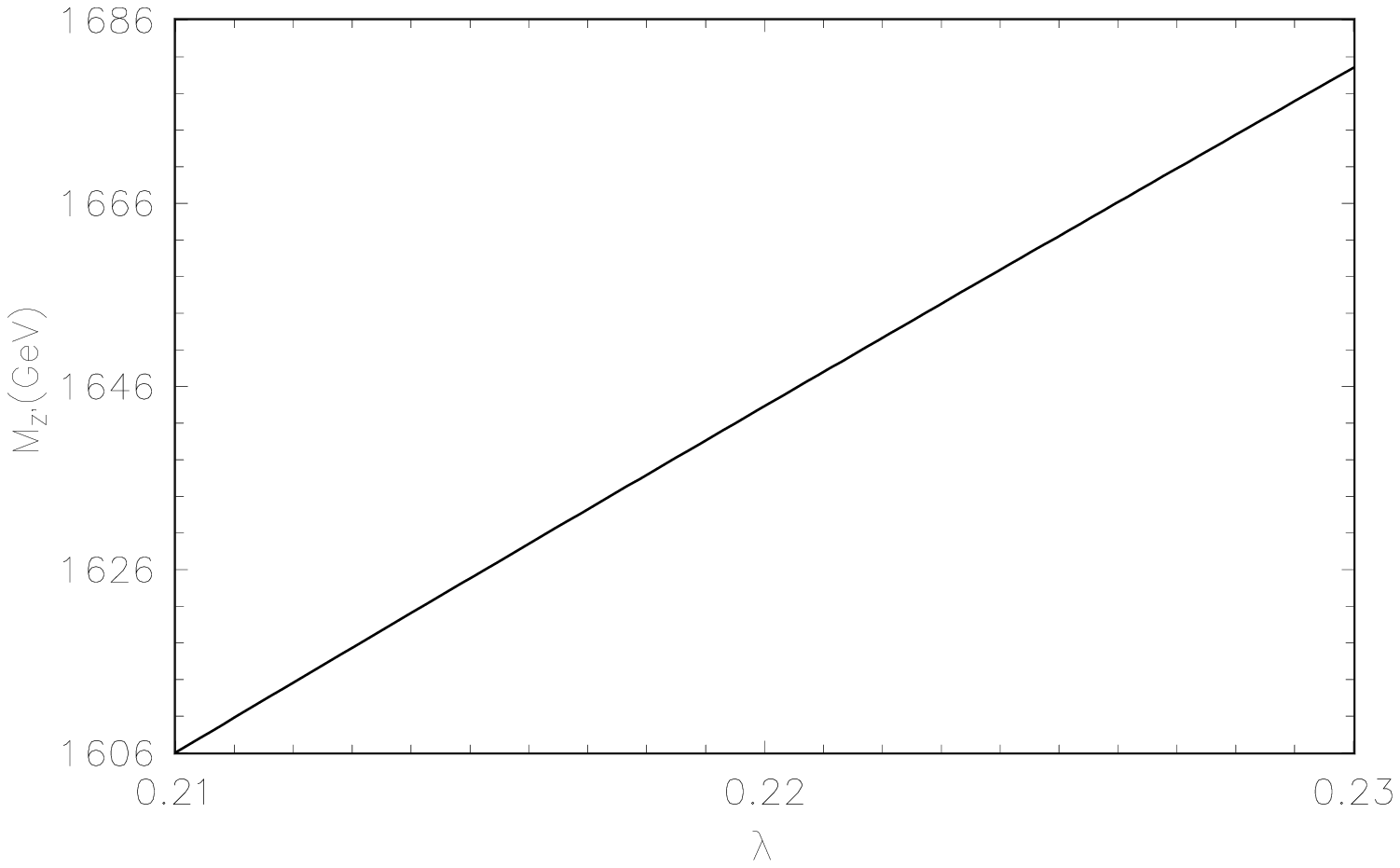}}
 \put(120,-95){\huge Fig.2}
 \end{picture}
 \end{center}
 \end{figure}
 \end{document}